\documentclass[aps,pra,twocolumn,showpacs]{revtex4-1} 
\usepackage{graphicx,amssymb,amsfonts,amsmath}

\usepackage{hyperref,color}

\usepackage{amsmath,mathtools}
\usepackage{bbold}

\newcommand{\be}{\begin{eqnarray}}
\newcommand{\ee}{\end{eqnarray}}

\newcommand{\n}{\nonumber \\}
\newcommand{\bp}{\begin{pmatrix}}
\newcommand{\ep}{\end{pmatrix}}            
\newcommand{\eq}[1]{Eq.~(\ref{#1})}   
\newcommand{\balpha}{\boldsymbol \alpha}            

\usepackage[utf8]{inputenc} 
\usepackage{graphicx}
\begin{document}

\title{Stochastic Differential Equations for Quantum Dynamics \\of Spin-Boson Networks}

\date{October 12, 2014}

\author{Stephan Mandt}
\affiliation{Department of Physics, Princeton University, Princeton, NJ 08544, USA \\
		Department of Computer Science, Columbia University, NY 10027, USA}
\author{Darius Sadri}
\affiliation{Department of Electrical Engineering, Princeton University, Princeton, NJ 08544, USA}
\author{Andrew A. Houck}
\affiliation{Department of Electrical Engineering, Princeton University, Princeton, NJ 08544, USA}
\author{Hakan E. T\"{u}reci}
\affiliation{Department of Electrical Engineering, Princeton University, Princeton, NJ 08544, USA}

\date{\today}

\begin{abstract}
The quantum dynamics of open many-body systems poses a challenge for computational approaches. 
Here we develop a stochastic scheme based on the positive $P$ phase-space representation to study the nonequilibrium dynamics of coupled spin-boson networks that are driven and dissipative.
Such problems are at the forefront of experimental research in cavity and solid state realizations of quantum optics, as well as cold atom physics, trapped ions and superconducting circuits.
We demonstrate and test our method on a driven, dissipative two-site system, each site involving a spin coupled to a photonic mode, with photons hopping between the sites,
where we find good agreement with Monte Carlo Wavefunction simulations.
In addition to numerically reproducing features recently observed in an experiment
\cite{PhysRevX.4.031043},
we also predict a novel steady state quantum dynamical phase transition for an asymmetric configuration of drive and dissipation.
\end{abstract}

\maketitle


\section{Introduction}
Our understanding of any physical system inevitably relies on our ability to subdivide the object of study into a
\textquotedblleft system\textquotedblright \ and a \textquotedblleft bath\textquotedblright.  In this sense, open quantum many-body systems are ubiquitous in nature and in practical applications. A wide range of theoretical and computational techniques have been developed in condensed matter physics to study many-body systems that equilibriate due to the coupling to their environment and can be described by equilibrium statistical mechanics in the long-time limit. In recent years attention has shifted to nonequilibrium many-body systems. For example their time evolution after a
quench~\cite{1742-5468-2007-06-P06008,RevModPhys.83.863,PhysRevLett.95.105701,PhysRevB.81.012303},
where the question of whether
and by what mechanism they thermalize~\cite{Rigol2008,Trotzky2012,Langen2013,ManyBodyLocalization,PhysRevA.88.043643} has become a focus of study.
Also of great interest are driven systems; 
applications range from the dynamics of
ultra-cold atoms~\cite{RevModPhys.80.885,Bloch2002,LewensteinBook,PhysRevLett.105.220405,Brantut08112013,Brantut31082012,PhysRevLett.104.080401},
trapped ions~\cite{PhysRevLett.112.023603},
coupled light-matter systems~\cite{Houck2012,ANDP:ANDP201200261,PhysRevX.4.031043,PhysRevLett.108.043003,PhysRevA.90.023820,Hur20082208,PhysRevA.90.023820},
transport problems \cite{Schneider2012,PhysRevLett.106.250602,PhysRevLett.108.086401,PhysRevB.86.085127},
and simulated quantum annealing~\cite{Boixo2014}. 
The need to account for quantum coherence that may be long-ranged in the presence of external forcing and dissipation is an intriguing problem that calls for the re-evaluation and extension of established techniques developed originally for near-equilibrium correlated systems. In recent years we have seen the further development of a number of such powerful computational techniques: exact diagonalization and density matrix renormalization group methods~\cite{RevModPhys.77.259,:/content/aip/proceeding/aipcp/10.1063/1.2080349},
nonequilibrium versions of dynamical mean field theory~\cite{RevModPhys.86.779},
application of Bethe ansatz techniques to nonequilibrium quantum transport in impurity models~\cite{PhysRevLett.96.216802},
and various Quantum Monte Carlo algorithms, among them the
continuous-time Monte Carlo algorithm (again for quantum impurity models)~\cite{RevModPhys.83.349}\footnote{The nondeterministic polynomial (NP) hard nature of the sign problem for fermions and frustrated magnets~\cite{Troyer2005} makes the general study of such systems hard, although for certain important transitions clever approaches can remove this
difficulty~\cite{Berg2012}.}.
Nonequilibrium quantum dynamics on the other hand has been a sine qua non of quantum optics and laser physics. In particular, an arsenal of powerful techniques have been developed in the field of Cavity QED to deal with nonequilibrium dynamics of
open quantum systems
~\cite{breuer2007theory,h1985light,gardiner2004quantum,Molmer:93,carmichael1999statistical,carmichael2009statistical}.
Master equation methods, methods based on Heisenberg-Langevin equations of motion, and Monte Carlo Wavefunction (MCW) approaches are by construction ideally suited to study dynamics of open quantum systems. With the experimental progress in Cavity QED in atomic, semiconductor and superconducting circuit systems  the attention has recently been drawn towards exploration of quantum many-body phenomena in extended light-matter systems. In particular lattices or networks of cavity QED
systems~\cite{Greentree2006,Hartmann2006,PhysRevA.76.031805,Koch2009,PhysRevA.82.043811,PhysRevA.82.052311,PhysRevA.86.023837,Houck2012,ANDP:ANDP201200261}
present a challenge to established techniques due to the exponential proliferation of the Hilbert space with the system size. For one dimensional systems DMRG-based approaches
~\cite{PhysRevLett.93.207205,PhysRevB.78.155117,1742-5468-2009-02-P02035,PhysRevLett.102.057202,PhysRevA.88.063835}
have been extended to study the dynamics of the density matrix of open quantum systems, but these rely on reduced dimensionality and certain constraints in the generation of entanglement during the evolution of the open system. There is a clear need
for advancing computational approaches that are more immune to the exponential growth problem and which scale more favorably with system size. Phase-space representations of quantum mechanics possibly offer such an approach, which we explore in this paper.  


Our goal in this paper is to develop a novel stochastic approach to study the dynamics of driven and dissipative systems involving spins and bosons, such as cavity QED systems.
We generalize the positive $P-$representation of quantum mechanics to model the dynamics of an interconnected network of spins and bosons coupled linearly to bosonic quantum baths. Phase-space representations have been employed in the past to study purely bosonic open systems 
~\cite{0305-4470-39-5-010,0305-4470-39-11-011},
but there is little work on spin systems and none that we know of for open spin-boson networks. Barry and Drummond~\cite{PhysRevA.78.052108} used the positive $P-$representation for spins to simulate
equilibrium thermodynamic properties of the quantum Ising model. Ng and S{\o}rensen~\cite{1751-8121-44-6-065305} used the mapping to Schwinger Bosons to derive a positive $P-$representation for a spin $1/2$ system. Closest to our approach are Refs.~\cite{PhysRevLett.108.073601,1367-2630-14-7-073011} which employ spin coherent states to derive a Fokker-Planck equation for the $Q-$function of the single-site Dicke model. In fact the present work is originally inspired by this work, to go beyond the Fokker-Planck level (which is numerically infeasible to solve) and develop a stochastic description, turning a formal identity into a numerical method. 
To this end a different representation is needed, as the $Q-$function does not possess a positive semi-definite diffusion matrix.

As we will present in some detail, we use a combination of bosonic and spin coherent states to map a quantum master equation to a Fokker-Planck equation, and in a second step, onto a stochastic differential equation which can be simulated efficiently. Notably, the latter step is only possible if the corresponding diffusion matrix in the Fokker-Planck equation is positive semi-definite. This is guaranteed in
the positive $P-$representation~\cite{0305-4470-13-7-018,0305-4470-39-11-011,PhysRevA.55.3014}.
To evaluate the effectiveness and accuracy of our computational approach, we analyze in detail a two-site system - a dimer - each site of which features a photonic mode coupled to a local spin
(this system has been recently studied in a circuit QED setup~\cite{PhysRevX.4.031043,PhysRevB.82.100507}).
We make a numerical comparison of our approach to the  Monte Carlo Wavefunction technique for spin
values accessible to the latter.
We refer to this system as the \emph{Dicke dimer} when the spins are taken to be large,
a limit which lies beyond the capability of the  Monte Carlo Wavefunction approach, but as we demonstrate is accessible in this new approach.
We stress that it can be applied to more complicated network geometries. 

Our paper is organized as follows: In section~\ref{sec:model}, we specify the quantum model that we
use as an example to demonstrate our formalism.
The first step in our mapping, the derivation of a Fokker-Planck equation from a quantum master equation
proceeds via the introduction of bosonic and spin coherent states, and is presented in ~\ref{sec:Fokker-Planck}.
In a second step, we map the Fokker-Planck equation to a stochastic differential equation in section~\ref{sec:SDE}.
The method is tested on a physical model in section.~\ref{sec:simul}. Finally, we summarize and discuss our results in section~\ref{sec:summary}.

\section{Spin-Boson Networks}
\label{sec:model}
A broad class of models in quantum optics and quantum information theory falls into the class of the following 
network-type, involving spins and bosons:
\be
\hat{H} & = & \sum_i \hat{H}_i \; + \; \hat{H}_{\rm kin}, \n
\hat{H}_{\rm kin} &=& -\sum_{ij} (J^{\rm spin}_{ij} \hat{S}_i^z\hat{S}_j^z + J^{\rm bos}_{ij} (\hat{a}^\dagger_i\hat{a}_j + \hat{a}^\dagger_j\hat{a}_i)) \label{eq:network1}
\ee
Parts of the Hamiltonian consists of a sum of local terms, describing e.g. external magnetic fields, on-site energies, coherent drives and interactions between bosons and spins.
The kinetic Hamiltonian couples different sites of the network.
A more specific class of models neglects the coupling of different spins, $J^{\rm spin}_{ij}=0$ , but still allows bosons to hop on the network. We also consider a specific type of on-site interactions:
\be
\hat{H} &=&\sum_{i} \hat{H}_i \,-\, \sum_{ij} J_{ij}(\hat{a}_i^\dagger \hat{a}_j + \hat{a}_j^\dagger \hat{a}_i) \, , \label{eq:network2}\\
\hat{H}_i &=& \omega_c \hat{a}_i^\dagger \hat{a}_i +  \omega_s \hat{S}^z_i +
\frac{g}{\sqrt{s}} 
\left(\hat{a}_i^\dagger \hat{S}^{-}_i + \hat{a}_i \hat{S}^+_i \right) + \nonumber \\
&& i f_i(\hat{a}_i - \hat{a}_i^\dagger) \, . \nonumber
\ee
For spin $1/2$ the single site system is known as the Jaynes-Cummings model, and it plays a prominent role in cavity and circuit QED systems, but appears also in other contexts.
We consider its generalization to a network, where the spin sizes are arbitrary.
We will refer to this model as the \textit{Dicke network}. 
It is characterized by a matrix of photon hopping amplitudes $J_{ij}$, a cavity frequency $\omega_c$, a spin frequency $\omega_s$, a matter-light coupling of strength $g$,
and coherent drive amplitudes $f_i$ at each site. Furthermore, the quantum number $s$ specifies the spin representation. 
Note that our model does not contain any counter-rotating terms.
Hence, the isolated Dicke network ($f_i=0$) conserves the total excitation number
$\sum_{i} \hat{N}_i + \hat{S}^z_i$, where $\hat{N}_{i}$ denotes the
photon number on site $i$, and $\hat{S}^z_{i}$ are the corresponding $z$ components of the spin.

Models ~(\ref{eq:network1}) and~(\ref{eq:network2}) are tractable with our approach that we present in this paper.
We focus on the dynamics of model~(\ref{eq:network2}), including possibly a coupling of the system to a bath. 
We therefore use the Lindblad master equation
\be
\partial_t \hat{\rho} &=& L[\hat{\rho}] \, , \label{eq:master_all} \\
L[\hat{\rho}] &=& -i[\hat{H}, \hat{\rho}] +
\sum_{i} ({\cal L}_i^a[\hat{\rho}] + {\cal L}_i^{S}[\hat{\rho}]) \label{eq:master} \, .
\nonumber
\ee
Dissipation in this master equation is described by the Lindblad superoperators
for the spins ${\cal L}^S$ and photons ${\cal L}^a$, describing weak coupling to a bath (we omit the site indices $i$ to keep the notation simple):
\be
{\cal L}^a[\hat{\rho}] &=& \frac{\kappa}{2}(\bar{n}+1)\left(
2\hat{a} \hat{\rho} \hat{a}^\dagger-
\hat{a}^\dagger \hat{a} \hat{\rho} -
\hat{\rho} \hat{a}^\dagger \hat{a} \right) \\
&& + \frac{\kappa}{2}\bar{n}\left(
2\hat{a}^\dagger \hat{\rho} \hat{a} -
\hat{a} \hat{a}^\dagger \hat{\rho} -
\hat{\rho} \hat{a} \hat{a}^\dagger \right) , \nonumber
\ee
\be
{\cal L}^{S}[\hat{\rho}] &=&
\frac{\gamma}{2}(\bar{n}+1)\left(
2 \hat{S}_ - \hat{\rho} \hat{S}_+ -
\hat{S}_ + \hat{S}_ - \hat{\rho} -
\hat{\rho} \hat{S}_ + \hat{S}_- \right) \\
& & + \frac{\gamma}{2}\bar{n}\left(
2 \hat{S}_+ \hat{\rho} \hat{S}_- -
\hat{S}_ - \hat{S}_+ \hat{\rho} -
\hat{\rho} \hat{S}_ - \hat{S}_+ \right). \nonumber
\ee
Here, the constants $\kappa$ and $\gamma$ specify the decay rate of photons from each cavity,
and the spontaneous decay rate of each spin.
$\bar{n}$ is the number of photons in the thermal bath and is a measure of temperature (with $\bar{n}=0$ for zero temperature)~\cite{gardiner2004quantum,carmichael1999statistical,carmichael2009statistical}.

Simulating this master equation numerically becomes intractable for large photon numbers, spins, and/or large networks.
As $L[\hat{\rho}]$ is a linear operator on \textit{density matrices}, the computational complexity grows quadratically in the Hilbert space dimension, which itself would grow exponentially with network size.
We thus aim for a different representation of the problem.

\section{Fokker-Planck equation}
\label{sec:Fokker-Planck}
\subsection{Coherent states and the positive $P$-representation}
\label{sec:coherent}

In this section we extend
the positive $P$-representation~\cite{gardiner2004quantum,carmichael1999statistical,carmichael2009statistical} to situations
involving spin.
The positive $P$-representation makes use of the basis of coherent states, which for bosons are
eigenstates of the annihilation operator $\hat{a}$ with eigenvalue $\alpha$,
\be
| \alpha \rangle = e^{-|\alpha|^2/2 + \alpha \hat{a}^\dagger}|{\rm vac}\rangle \, ,
\ee
and $|{\rm vac}\rangle$ denoting the vacuum state for the bosons. 
The spin coherent states~\cite{0022-3689-4-3-009,PhysRevA.6.2211,Stone1989557}
for the spin $s$ representation of the $su(2)$ spin algebra will be defined as
\be
|z\rangle = \frac{e^{z \hat{S}_+}}{(1+|z|^2)^s} |s,-s\rangle \, ,
\ee
where $\hat{S}_+$ is the raising operator the algebra.
Both state labels $\alpha$ and $z$ are complex valued.
Note that in our convention, the spin state $|z=0\rangle$ corresponds to the lowest weight
(\textquotedblleft spin down\textquotedblright) state.
We construct  the following operators:
\be
\hat{\Lambda}_a(\alpha,\beta) = \frac{|\alpha \rangle \langle \beta^*|}{\langle \beta^*|\alpha\rangle} =
\frac{e^{\alpha \hat{a}^\dagger}|\rm vac\rangle\langle \rm vac | e^{\beta \hat{a}}}{e^{\alpha\beta}} \, .
\ee
We similarly define for the spins
\be
\hat{\Lambda}_{S}(z,w) =
\frac{|z \rangle \langle w^*|}{\langle w^*|z\rangle} =
\frac{e^{z \hat{S}_+}|s,-s\rangle\langle s,-s| e^{w \hat{S}_-}}{(1+wz)^{2s}} \label{eq:lambdasigma} \, .
\ee
The normalization of $\hat{\Lambda}_S$ comes from the overlap of two spin coherent 
states~\cite{0022-3689-4-3-009,PhysRevA.6.2211,Stone1989557}.
We now introduce a container variable for the complex numbers that specify sets of operators,
$\hat{\Lambda}_a$ and $\hat{\Lambda}_S$, for the network:
\be 
\balpha = (\alpha_1,\beta_1,z_1,w_1, \cdots, \alpha_n,\beta_n,z_n,w_n) \, .
\ee 
where $n$ is the size of the network.
Combining spin and bosonic degrees of freedom, we then define the following operator which acts on the
full many-body Hilbert space of the network:
\be
\hat{\Lambda}(\balpha)&=&\prod_{i} \, \hat{\Lambda}_a(\alpha_i,\beta_i)\otimes \hat{\Lambda}_S(z_i,w_i) \label{eq:defineLambda} \, .
\ee
The density operator of the system can now be expanded in
our generalized positive $P$-representation as follows:
\be
\hat{\rho}(t) = \int d\balpha \: P(\balpha,t)\: \hat{\Lambda}(\balpha)\, .
\ee
In the above, we defined the integration measure
$d\balpha = \prod_{i} \, d^2\alpha_i \, d^2\beta_i \, d^2z_i \, d^2 w_i$.
As can be easily verified, normal-ordered bosonic operator expectation values in the positive $P$-representation are calculated according to
\be
\langle \hat{a_i}^{\dagger n} \hat{a_j}^m \rangle &=& \int d\balpha \; P(\balpha,t)\, \beta_i^n \alpha_j^m \\
& \approx & \frac{1}{N_s}\sum_{l=1}^{N_s}\beta_{l,i}^n(t) \alpha_{l,j}^m(t)
\ee
The second line gives an approximation to the expectation value for the case where only $N_s$ samples $ \alpha_l(t),\beta_l(t)$, from the positive $P$-function
are available (in terms of solutions to an equivalent stochastic differential equation to be derived
below).
An example of an expectation value involving spin is given in appendix
\ref{appendix:spin_expectation_values}.

\subsection{Mapping to a Fokker-Planck equation}
\label{sec:mapping}

Having introduced coherent states and $P$-functions, it is worth outlining the general strategy for deriving a Fokker-Planck equation.
Consider a general master equation of the form 
\be
\dot{\hat{\rho}} = L [\hat{\rho}] \, ,
\ee
where $L$ is an arbitrary Liouvillian or  Lindblad operator. As we will show in the next paragraph,
the previously introduced operators $\hat{\Lambda}$ allow us to convert second-quantized operators into differential ones. As a consequence,
\be
L[\hat{\Lambda}(\balpha)] &=& {\cal D}^L(\balpha)\hat{\Lambda}(\balpha) \label{eq:Lcorrespondence}
\ee
where ${\cal D}^L(\balpha)$ is a representation of $L$ in terms of derivatives with respect to $\balpha$. 
It turns out that this differential operator consists only of first and second order derivatives and can therefore be written as
(Einstein summation convention over site indices is implied)
\be
{\cal D}^L(\balpha) \;=\; -A_i(\balpha)\partial_{\balpha_i} + \frac{1}{2} D_{ij}(\balpha)\partial_{\balpha_i} \partial_{\balpha_j} \label{eq:formofD} \, .
\ee
The dependence of the vector $A$ and the matrix $D$ on $\balpha$ will be determined shortly.
Using this relation (without further specifying ${\cal D}^L(\balpha)$ at this point), we can derive 
a Fokker-Planck equation from the Lindblad master equation as follows. 
We first note that
\be
L[\hat{\rho}]  &=& \int d\balpha \, P(\balpha,t) L\left[\hat{\Lambda}(\balpha)\right]  \stackrel{(\ref{eq:Lcorrespondence})}{=} \int d\balpha \, P(\balpha,t) {\cal D}^L \hat{\Lambda}(\balpha) \, . \nonumber
\ee 
The derivatives can be transferred from $\hat{\Lambda}$ to $P$ by partial integration. We use eq.~(\ref{eq:formofD}) to find
\be
\int d\balpha \, && \partial_t P(\balpha,t) \hat{\Lambda}(\balpha) = \\
&& \int d\balpha \, \hat{\Lambda}(\balpha)  \left(A_i \partial_{\balpha_i} + \frac{1}{2}  D_{ij}\partial_{\balpha_i} \partial_{\balpha_j} \right)P(\balpha,t)\nonumber \, ,
\ee
from which we conclude that 
\be
\label{eq:formal-FP}
\partial_t P(\balpha,t) = \left(A_i \partial_{\balpha_i} + \frac{1}{2}  D_{ij}\partial_{\balpha_i} \partial_{\balpha_j} \right)P(\balpha,t) \, .
\ee
This is the Fokker-Planck equation. The goal for the remainder of this section is to calculate $A_i(\balpha)$ and $D_{ij}(\balpha)$.
Note that the better known $P$-representation is simply obtained from the positive $P$-representation by substituting $\beta \rightarrow \alpha^*$, which 
enforces the variables $\alpha$ and $\beta$ to be complex conjugates.  In the positive $P$-representation, the presence of quantum noise violates this conjugacy relation;
for more details we refer the reader to~\cite{gardiner2004quantum}.

Eq.~(\ref{eq:Lcorrespondence}) provides a prescription for finding the differential operator ${\cal D}^L$,
by evaluating the action of the Lindblad operator appearing in eq.~(\ref{eq:master_all}) on $\hat{\Lambda}$:
\be
L[\hat{\Lambda}] &=& \; i \sum_{ij }J_{ij} \, [\hat{a}_i^\dagger \hat{a}_j + \hat{a}_j^\dagger \hat{a}_i, \hat{\Lambda}]   \label{eq:LonLambda} \\
&& + \sum_{i} \left(-i [\hat{H}_i,\hat{\Lambda}]+ {\cal L}_i^a[\hat{\Lambda}] + {\cal L}_i^{S}[\hat{\Lambda}] \right)
\, . \nonumber
\ee
In order to proceed, we need the action of the second-quantized operators appearing
in the Hamiltonian on $\hat{\Lambda}$,  (see Ref.~\cite{gardiner2004quantum}  for a pedagogical introduction to the positive P representation).
We list and derive those identities in appendix~\ref{appendix:spin_derivation}. 
Note that each creation or annihilation operator corresponds 
to a \textit{first order} differential operator. Commutation relations reflect themselves in the
non-commutativity of these differential operators.
Furthermore, the master equation
contains only products up to second order in bosonic creation and annihilation operators and spin raising and lowering operators.
This guarantees that  the resulting partial differential equation is necessarily 
of second order in $\balpha$ and first order in time, and hence of Fokker-Planck type \footnote{We stress that there are a variety of alternative approaches to deriving a Fokker-Planck equation,
for example by choosing a different phase-space representation ($Q$, $P$, Wigner,  etc.), a different normalization
for the operators $\hat{\Lambda}$ as in~\cite{PhysRevA.78.052108}, or by choosing a different quantization axis for the spin coherent states}.

We first simplify the contributions arising from the local Hamiltonians $\hat{H}_i$ (derivations are presented in the appendices).
Note that all complex fields $\alpha,\beta,z,w$ carry a site index. In order not to overload the notation at this point, we omit these indices and instead indicate the site index on the brackets: 
\begin{widetext}
\begin{align}
\Big[ \hat{H}_i,\hat{\Lambda}(\balpha) \Big] &= \nonumber \\
&\Bigg[ \left(\omega_c \alpha + f+ 2g \sqrt{s} \frac{z}{1+wz}\right)\partial_\alpha - \left(\omega_c\beta + f + 2g\sqrt{s}\frac{w}{1+wz}\right)\partial_\beta \nonumber \\
& \ \ \ +  \left(\frac{g}{\sqrt{s}}\left(\alpha-\beta z^2\right)+\omega_s z\right)\partial_z - \left(\frac{g}{\sqrt{s}}
\left(\beta-\alpha w^2\right)+\omega_s w\right)\partial_w
+ \frac{g}{\sqrt{s}}
\left(w^2 \partial_w \partial_\beta - z^2 \partial_\alpha \partial_z\right)
\Bigg]_i \hat{\Lambda}(\balpha) \, .
\end{align}
\end{widetext}
We next compute the kinetic term for the network, being the only term coupling the different sites:
\be
\Big[ (\hat{a}_i^\dagger \hat{a}_j &+& \hat{a}_j^\dagger \hat{a}_i) ,\hat{\Lambda}(\balpha) \Big]
= \\
&& \left( \alpha_j\partial_{\alpha_i} + \alpha_i\partial_{\alpha_j}
- \beta_j\partial_{\beta_i} - \beta_i\partial_{\beta_j} \right) \hat{\Lambda}(\balpha) \, . \nonumber
\ee
Finally, we calculate the dissipators, starting with the photons. After a straightforward calculation,
making use of \eqref{photon_identities}, we find
\be
{\cal L}^{a}_{\rm}[\hat{\Lambda}(\balpha)] &=&
\frac{\kappa}{2} \left( - \alpha \partial_\alpha - \beta \partial_\beta + 2 \bar{n} \,\partial_\alpha \partial_\beta \right) \hat{\Lambda}(\balpha) \, .
\ee
Notably, all second-order derivatives are proportional to $\bar{n}$. According to the Feynman-Kac relation, second order derivatives correspond to noise terms
in a stochastic description. As $\bar{n}(T=0)=0$, there is no quantum noise associated to cavity loss at zero temperature.

A much longer calculation, shown in the appendix, results in the following contributions from the spin dissipators:
\be
 {\cal L}^{S}  [\hat{\Lambda}(\balpha) ] &=&  \Big[-\frac{\gamma}{2}(2\bar{n}+1)(z^2 \partial_z^2 + w^2 \partial_w^2) \\
&& + \gamma \left(\bar{n} + \left(\bar{n}+1 \right) z^2 w^2 \right) \partial_z \partial_w
\nonumber \\
&& +  \left( \gamma(- z\,s + \frac{\bar{n}}{2} z) \partial_z + (z \leftrightarrow w)\right) \Big]
\hat{\Lambda}(\balpha) \, . \nonumber  
\ee
The double arrow indicates an identical contribution on the last line where the roles of $z$ and $w$ are interchanged.
Interestingly, the spin dissipator \textit{does} contain quantum noise terms that are present even at zero temperature. 

Our goal will be to cleanly separate quantum from classical dynamics. To this end, we perform a transformation on the following variables:
\be
\tilde{\alpha}  =  \alpha  \sqrt{s} \, ,
\ \ \tilde{\beta}  =  \beta \sqrt{s} \, ,
\ \ \tilde{\bar{n}} = \bar{n} s \, ,
\ \ \tilde{f} = f \sqrt{s} \, .
\ee
However, in order to keep the notation clean, we omit the tildes in the remainder of the paper. 
The third transformation is needed as photon densities scale as $n\sim \alpha \beta$. Intuitively , as the field amplitude gets scaled, the photon
density of the bath and the external drive have to be scaled up as well (otherwise the bath temperature would be effectively lowered). After the transformation, the Hamiltonian contribution
to the Fokker-Planck equation is independent of $s$, and hence it has a well-defined limit for $s \rightarrow \infty$.
In contrast, the second-order differential operators that arise due to the interaction $g$ are proportional to $s^{-1}$
and therefore vanish in this limit. This means that quantum noise vanishes in the classical limit of large spin, as intuition would suggest~\cite{PhysRevLett.108.073601}.


\subsection{Drift vector and diffusion matrix}
We are now ready to collect all terms and specify the drift vector $A$ and the diffusion matrix $D$ in the Fokker-Planck equation~(\ref{eq:formal-FP}).

The vector A has $4$ complex entries for each network site. The 4 entries corresponding to site $i$ are given by
\be
\label{eq:determin_A}
A_1 (\balpha) &=& i \left(\omega_c \alpha_i - \sum_{j\neq i} J_{ij} \alpha_j+ f_i(t) + \frac{2g z_i}{1+w_iz_i} \right)-\frac{\kappa}{2}\alpha_i \, , \nonumber \\
A_2 (\balpha)&=& -i\left(\omega_c \beta_i - \sum_{j\neq i} J_{ij} \beta+ f_i(t) + \frac{2gw_i}{1+w_iz_i} \right)-\frac{\kappa}{2}\beta_i \, , \nonumber \\
A_3 (\balpha) &=& i g \left(\alpha_i - \beta_i z_i^2\right)+i\omega_s z_i - \gamma \left(1 -
\frac{\bar{n}}{2}\right)
z_i \, , \\ 
A_4 (\balpha) &=& -i g \left(\beta_i - \alpha_i w_i^2\right) -i \omega_s w_i - \gamma \left(1 -
\frac{\bar{n}}{2}\right) w_i \nonumber \, .
\ee
As we shall explain below in more detail, these are the right hand sides of the classical equations of motion for the variables $\alpha_i, \beta_i, z_i$ and $w_i$, respectively.

We now move on to the diffusion matrix $D$. Note that it is block-diagonal in the space of network sites;
therefore we will focus on a single site, omitting the site indices.
For later convenience, we will separate it in the following way
\be
\label{eq:decomp}
D
&=& D^{(1)} + D^{(2)} + D^{(3)} \\[.2cm]
&=&
\begin{pmatrix} 
D_\kappa &0\\ 0 & 0
\end{pmatrix} 
\, + \,
\begin{pmatrix}
0 &D_g\\ D_g & 0
\end{pmatrix}
 \, + \,
\begin{pmatrix}
0 &0\\ 0 & D_\gamma
\end{pmatrix}
\, . \nonumber
\ee
The non-zero entries are themselves $2 \times 2$ matrices
\be
D_\kappa  &=&   \bar{n}\,\kappa 
\begin{pmatrix}
0 & 1 \\ 1 & 0
\end{pmatrix} \, , \\
D_g  &=&  \frac{i g}{s}
\begin{pmatrix}
 -z^2&0 \\ 0 & w^2 
\end{pmatrix} \, , \nonumber \\
D_\gamma  &=& \gamma
\begin{pmatrix}
(2 \bar{n}+1)z^2 &  \bar{n}+( \bar{n}+1)z^2w^2 \\  \bar{n}+( \bar{n}+1)z^2w^2 & (2 \bar{n}+1)w^2 
\end{pmatrix} \, . \nonumber
\ee
These matrices are proportional to the parameters $\kappa$, $g$ and $\gamma$, respectively,
indicating three distinct sources of noise, namely a quantum noise contribution due to $g$, a purely thermal noise arising from $\kappa \bar{n}$,
and a noise contribution from the spin, associated with the spontaneous emission at rate $\gamma$.
In contrast to the other noise contributions, $D^{(2)}$ couples the photon and spin sectors.
Also note that it is proportional to $1/s$, and hence vanishes in the classical limit.

\section{Stochastic differential equations}
\label{sec:SDE}

In order for our approach to provide an efficient basis for numerical simulation,
we furthermore map the Fokker-Planck equation we have derived
onto a set of stochastic differential equations.
A necessary requirement for this step to be possible is that the diffusion matrix be positive semi-definite.
If this requirement is not met, then the diffusion matrix possesses 
contracting directions, and it is impossible to model contracting densities in terms
of random walks.
This requirement is precisely why we have chosen to work in the positive $P$-representation, 
as then the positivity of the diffusion matrix is
assured~\cite{gardiner2004quantum,carmichael1999statistical,carmichael2009statistical}.

It follows from the standard theory of stochastic calculus~\cite{carmichael1999statistical,carmichael2009statistical} that the Ito stochastic differential equations equivalent
to the Fokker-Planck equations~\eqref{eq:formal-FP} are given by
\be
d\balpha &=& A(\balpha) dt + d\xi(\balpha,t) \label{eq:stoch_diff_general} \\
\langle d\xi^\mu(t) d\xi^\nu(t')\rangle &=& D_{\mu \nu}(\balpha) \delta_{tt'}\,  \label{eq:stoch_diff_correlator} dt
\ee 
Hence, the deterministic evolution is described by the drift vector $A$, while the equal-time correlator of the noise in the four dimensional complex space is given by the diffusion matrix. We are now faced with the challenge of designing a noise such it satisfies this requirement. For the following discussion, we focus on a single network site and omit the site index (the diffusion matrix is block-diagonal).
An obvious way of creating such a noise term would be to define
\be
d\xi(\balpha,t) &=& B(\balpha,t) \, dW(t) \, , \n
D(\balpha) &=& B(\balpha,t)B(\balpha,t)^T \, , \n
\langle dW(t) dW^T(t')\rangle &=& \delta_{tt'}dt \,\mathbb{1}_4 \, ,
\ee
where $\mathbb{1}_4$ is the four dimensional identity matrix.
The noise is thus decomposed into the product of $B(\balpha,t)$ (the matrix square root of $D$), with $dW$,
a four dimensional vector of independent Wiener increments. 
However, as $D$ is a complex $4\times 4$  (or real $8 \times 8$) matrix for each network site, determining this matrix decomposition numerically at each infinitesimal time step would be computationally demanding.

Fortunately, we can use a trick to circumvent the need to perform such a time-dependent factorization,
making the numerical algorithms more efficient,
by using the explicit decomposition of $D$ given in eq.~\eqref{eq:decomp}. 
In the following, we will construct three infinitesimal noise vectors $d\xi_1(t), d\xi_2(t)$, and $d\xi_3(t)$, which are taken to be mutually uncorrelated, and
which satisfy for each $i=1,2,3$ (no summation over $i$)
\be
\label{eq:xi_i}
\langle d\xi_i^\mu(t) d\xi_i^\nu(t')\rangle &=& D^{(i)}_{\mu \nu}(\balpha) \delta_{tt'}\, dt \, ,
\ee
with the individual $D^{(i)}$ given in eq.~\eqref{eq:decomp}. We then define the total noise
\be
d\xi(t) = d\xi_1(t) + d\xi_2(t) + d\xi_3(t) \, .
\ee
As the $d\xi_i$ are mutually uncorrelated, it follows that 
\be
\langle d\xi^\mu(t) d\xi^\nu(t') \rangle &=& \sum_{i=1}^3\langle d\xi_i^\mu(t) d\xi_i^\nu(t') \rangle \\
&=& D_{\mu \nu} \delta_{tt'}\, dt \, , \nonumber
\ee
after using \eqref{eq:decomp} and \eqref{eq:xi_i}.
Consequently we have succeeded in generating a random noise vector with the desired correlator~(\ref{eq:stoch_diff_correlator}). It remains still to show how to construct
the individual $d\xi_i$'s. This is, however, easy. 
For this purpose, we construct the matrix square roots $B^{(i)}$ for the matrices $D^{(i)}$ such that  $D^{(i)} = B^{(i)}B^{(i),T}$. 
Due the structural simplicity of the matrices $D^{(i)}$ we find the analytic solutions
\be
\label{eq:root_diffusion_matrix}
B^{(1)} &=& \sqrt{\kappa \bar{n}} 
\frac{1}{\sqrt{2}}\begin{pmatrix}
1 & i & 0 & 0\\ 1 & -i & 0 & 0 \\ 0 & 0 & 0 & 0 \\0 & 0 & 0 & 0
\end{pmatrix} \, ,
 \\
B^{(2)}&=& \sqrt{\frac{i g}{s}} 
\frac{1}{\sqrt{2}}
\begin{pmatrix}
1 & 1 & 0 & 0\\
0 & 0  & 1 & -1 \\
-1 & 1 & 0 & 0 \\ 
0 & 0 & 1 & 1 
\end{pmatrix} 
\begin{pmatrix}
z & 0 & 0 & 0\n
0 & i z  & 0 & 0 \\
0 & 0 & w & 0 \\ 
0 & 0 & 0 & iw 
\end{pmatrix} \, . \nonumber 
\ee
A similar analytic formula for $B^{(3)}$ can be obtained from diagonalizing a $2\times2$ matrix, but in the following we will set the spontaneous emission rate $\gamma$ to zero, and so have no need for it.
We now define 
\be
 \label{eq:threenoises}
d\xi_i(t) &=& B^{(i)} dW^i(t) \quad \quad (i=1,2,3) \, ,
\ee
with each $dW^i(t)$ being a four dimensional vector of independent Wiener increments.  It follows that the individual 
noises $d\xi_i$ have the correlators (\ref{eq:xi_i}). This concludes our construction of the correlated noise.

We note here that our noise vector can be multiplied from the right by {\it any} complex orthogonal matrix, and would still satisfy eq.~(\ref{eq:stoch_diff_correlator}).
This observation is a realization of the stochastic gauge degree of freedom~\cite{1464-4266-5-3-359,1751-8121-44-6-065305}. 

\section{Numerical simulations}
\label{sec:simul}
\subsection*{Nonequilibrium Dicke-dimer}
We shall now apply our stochastic formalism to a physical test case, one involving
strong spin-photon interactions, and which has an additional spatial degree of freedom involving a kinetic energy term. For simplicity we will focus on the
dynamics of two coupled cavities (a dimer), each of which contains a spin
coupled to a single photonic mode. In circuit Quantum Electrodynamics (circuit QED),
each site would be realized 
in terms of a microwave field, coupled to a superconducting qubit, with the sites capacitively coupled to
allow photon hopping.
As the qubit can be interpreted as a spin $s=1/2$, each cavity is
described by a Jaynes-Cummings model.
A dimer of such cavities has recently been studied in an experiment~\cite{PhysRevX.4.031043}.
We consider a broader class of similar systems, allowing for arbitrary spin $s$.
In particular, we are interested in the scaling limit of large spin and photon numbers
and in the quantum to classical crossover. A corresponding dimer could  e.g. be realized by using many qubits per cavity that are all coupled to the same photonic mode.

To begin with, let us briefly review the main experimental findings.
In the experiment, one of the two cavities (cavity 1, which we will take to be the left cavity) is initially
populated with many photons.  The system is undriven and is let to evolve in time.
As both cavities are dissipative (with photon loss rate $\kappa$), the photon number
decreases monotonically with time.
The following physics is observed in the experiment.
At large photon numbers, photons are observed to undergo linear periodic
oscillations between the two cavities, while simultaneously exponentially decaying to the outside environment. 
However, as the photon number drops below a certain critical threshold, 
the oscillations are seen to cease as the system enters a macroscopic quantum self-trapped state (for details please refer to~\cite{PhysRevX.4.031043}).

A qualitative explanation of the physics is as follows: There are two competing time scales for the dimer
when observing the homodyne signal (equivalently, there is a competition between the on-site interaction energy, with scale set by the cavity-spin coupling $g$, and the kinetic energy, dictated by the hopping rate $J$).
The Josephson oscillations occur with period $t_J=1/2J$ when the photon number is above the
critical threshold and the system is in the delocalized phase.
The second time scale is the collapse and revival period associated with
the single site Jaynes-Cummings physics, the relevant time scale when the system has localized and
the tunneling disappears, wherein the two sites are effectively decoupled.
The localization transition is predicted to occur when these two time scales become comparable.
This matching argument has been supported by extensive numerical simulations for the spin $1/2$ dimer
using Monte Carlo Wavefunction simulations~\cite{PhysRevX.4.031043}.

We would like to analyze the quantum transition in the well-controlled semiclassical limit of large spin, going beyond the classical solution and taking into account
the impact of quantum fluctuations, as well as the effect of thermal noise. 
We will also give a theoretical explanation for the super-exponential decay of the homodyne signal
that has been observed in~\cite{PhysRevX.4.031043}. 

We test our  method on the example of a dissipative Dicke dimer. We are interested in two cases. 
First, we study the undriven lossy dimer, inspired by the experiment, where we prepare an initial state with a fixed number of photons in the left cavity
and study the dynamics. 
Second, we study the corresponding driven system. Here, we are mainly interested in the 
behavior of the tunneling current between the two sites, in steady state.
As we show, the driven system displays a dynamic quantum phase transition visible in the inter-cavity current 
upon varying the interaction strength.

\begin{figure}[h]
\begin{center}
\includegraphics[width = \linewidth]{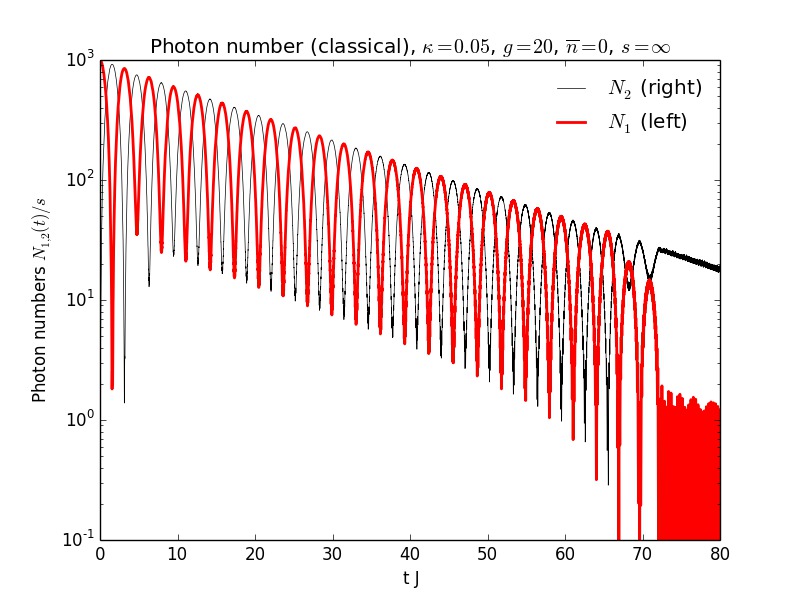}
\caption{Deterministic semiclassical equations, corresponding to the scaling limit $s=\infty$ with $\bar{n}=0$ (no thermal noise). 
The plot shows the numbers of photons in the first (bold, red) and second (thin, black) cavity, respectively. }
\label{fig:n_left}
\end{center}
\end{figure}

\subsection{Undriven dissipative Dicke dimer}
\paragraph*{\bf{Zero temperature, infinite spin.}}
We begin by modeling the classical dynamics of the dissipative Dicke dimer at $T=0$. Zero temperature
($\bar{n}=0$) in combination with infinite spin implies that all noise terms vanish. The
equations~\eqref{eq:stoch_diff_general}
are then completely deterministic, and the positive $P$-representation becomes
equivalent to the standard $P$-representation,
allowing for an alternative representation of the
stochastic differential equations in terms of the compact angular variables.
We have found the corresponding equations~\eqref{eq:SDE_spherical} to be more stable at long times in this new representation.
Simulation results for a decaying dimer are presented in figure~\ref{fig:n_left}. 
We observe a self-trapping transition setting in at a critical photon number, below which
the oscillations die out rapidly.
The dynamic equations in this limit are equivalent to the classical Maxwell-Bloch equations that have been studied earlier in this setup~\cite{PhysRevB.82.100507,PhysRevX.4.031043},
but whose derivation requires the uncontrolled assumption of the factorization of operator expectation values, whereas our deviation is fully controlled in the scaling limit.

\begin{figure}[h]
\begin{center}
\includegraphics[width = \linewidth]{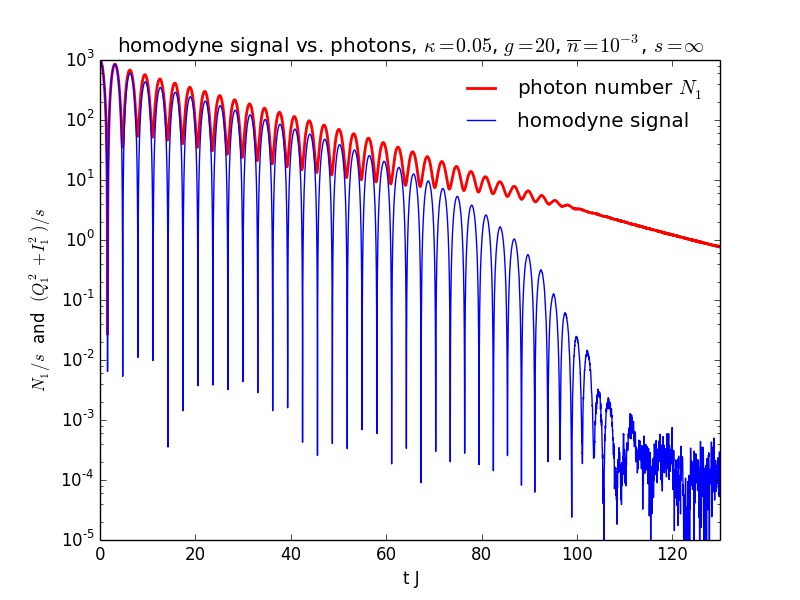}
\caption{Finite temperature, classical simulations for the self-trapping transition (infinite spin). We averaged over 10,000 stochastic trajectories. 
The plot shows particle number (red) and homodyne signal (blue). At the transition, the homodyne signal decays super-exponentially.}
\label{fig:homodyne}
\end{center}
\end{figure}

\paragraph*{\bf{Finite temperature, infinite spin.}}
Next, we explore the impact of thermal noise. To this end, we set  $\bar{n}$ to a finite, positive value,
simulating the coupling to an external photon bath with mean occupation $\bar{n}$. 

The fact that we have taken the spin $s \rightarrow \infty$ and the qubit relaxation rate
$\gamma \rightarrow 0$ implies that the only remaining noise term in eqs.~(\ref{eq:root_diffusion_matrix},\ref{eq:threenoises})
is $\xi_1(t)$, corresponding to the matrix $B^{(1)}$. This matrix has an interesting symmetry property: when multiplying on the right any real vector,
the resulting complex vector has just two entries which are always complex conjugates. As a consequence,
the random thermal noise acting on the variables $\alpha$ and $\beta$ preserves this conjugacy. The drift equations share the same property, and preserve conjugacy,
namely $A_1(\alpha) = A_2^*(\beta)$ for $\beta = \alpha^*$,
where $*$ denotes complex conjugation. It follows that $\alpha(t) = \beta^*(t)$ for all times and all stochastic trajectories.
This  mirrors the fact that the positive $P$-representation is equivalent to the ordinary $P$-representation
in the absence of interaction induced (quantum) noise.

Physically, coupling a thermal bath to our decaying cavity will induce two things: first, coherence will be destroyed over time, and second, the system's equilibrium
state (at least for small $g$) will not be the vacuum state but rather an incoherent photon state at mean photon number $\bar{n}$. To illustrate the loss of coherence,
we calculated the \textit{homodyne signal}
$h = \langle \hat{I}\rangle^2 +\langle \hat{Q}\rangle^2$,
where the quadratures are defined in terms of the creation and annihilation operators as
$\hat{I}=(1/2)(\hat{a}+\hat{a}^\dagger)$ and
$\hat{Q}=(i/2)(\hat{a}^\dagger-\hat{a})$.
This quantity was experimentally measured
in Ref.~\cite{PhysRevX.4.031043} in the closely related setup of a decaying Jaynes-Cummings dimer, i.e. for $s=1/2$. Note that for a perfectly coherent system where
$\langle \hat{a}^\dagger \hat{a}\rangle = \langle \hat{a}^\dagger \rangle\langle \hat{a}\rangle$, the homodyne signal measures the photon number. In the presence of some incoherence, however, it drops below
the photon number.
In Ref.~\cite{PhysRevX.4.031043}, the homodyne signal was seen to decay super-exponentially close to the self-trapping transition. Our simulations show a qualitatively similar behavior in 
figure~\ref{fig:homodyne}, where the homodyne signal (but not the photon number)
is seen to decay super-exponentially.
In the experiment, individual photons escape according to a Poisson process.
For each single trajectory in the ensemble average,
the photon number drops below the critical threshold at a random
time, the initial photon number determining the average time at which this occurs.
On approaching the transition, the oscillations become highly nonlinear, with a diverging
period (critical slowing down)~\cite{PhysRevB.82.100507}.
This results in a dephasing of the different trials within an ensemble.
Therefore averages of the homodyne signal die out faster than exponentially.
Hence, the quantum localization transition in~\cite{PhysRevX.4.031043}
possesses a classical analogue at large spin.

\begin{figure}
\begin{center}
\includegraphics[width = \linewidth]{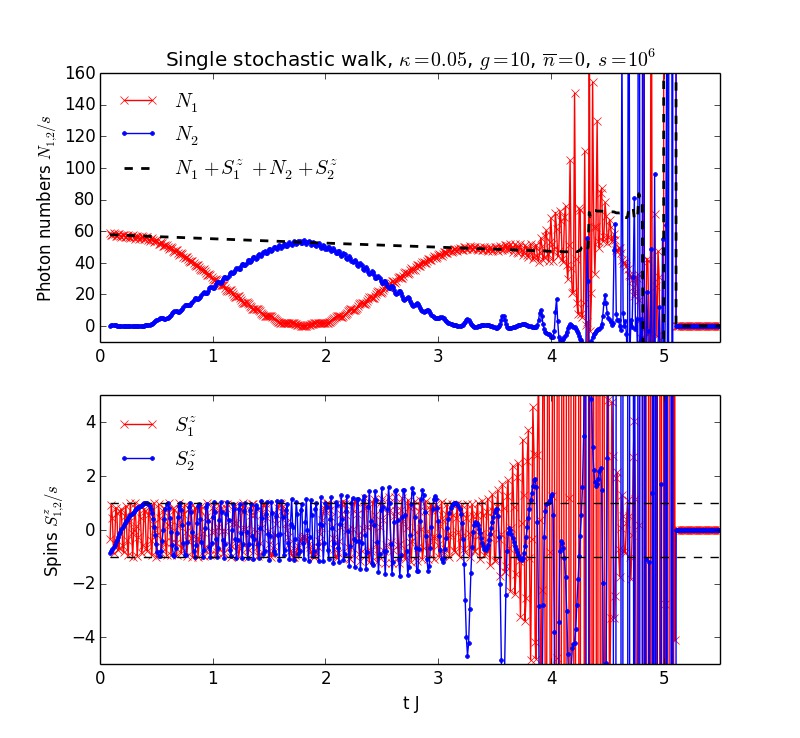}
\caption{Single stochastic trajectories for finite spin. Large spins (here $s=10^6$) allow us to simulate for long times and thus follow the dynamics into self-trapping in the quantum regime (it is important to note, however, that we are here only showing a single trajectory, not an ensemble average). The upper panel shows the photon numbers in the left (red crosses) and right (blue dots) cavities, respectively. The high frequency oscillations are Rabi oscillations, whose frequency depend on the photon number. They 
are also apparent in the lower panel, showing the $z$-components of the two spins (same color coding).
We also present the system's total excitation number
$\langle\hat{N}_1 + \hat{N}_2 + \hat{S}^z_1 + \hat{S}^z_2\rangle$ (black dashed line in upper panel), which is a constant of motion for the closed system,
but here slowly and smoothly decays due to the cavity loss. Jumps in this line indicate the breakdown of numerical reliability, here at  $t=4.3$.}
\label{fig:pos_P_single_traject}
\end{center}
\end{figure}

\paragraph*{\bf{Finite spin}}
In contrast to the semiclassical limit of infinite spin, finite spin simulations require the full machinery of the positive $P$-representation.
In particular, the emergent quantum noise at finite spin violates the
conjugacy relation between $\alpha$ and $\beta$ (and also $z$ and $w$). Hence, all four complex coordinates evolve according
to their individual dynamics, and all are subject to individual (yet correlated) sources of noise. 
This fact has counterintuitive consequences. Let us consider for example of the photon density in a given cavity. 
An individual stochastic trajectory in the $P$-representation necessarily has positive photon numbers $n\sim \alpha(t) \alpha^*(t) \in {\mathbb R_+}$.
In positive $P$-representation, individual runs have generally complex contributions $n\sim \alpha(t)\beta(t) \in {\mathbb C}$. It is therefore
important to keep in mind that only \textit{averaged quantities} have a physical meaning. Similarly, the $z-$component of the
spins in the $P$-representation are given by $\hat{S}_z \sim  \frac{1-zz^*}{1+zz^*} \in [-1,1]$. In contrast, the corresponding contribution
in positive $P$-representation reads $\hat{S}_z \sim  \frac{1-zw}{1+zw} \in {\mathbb C}$. This behavior is seen
for example in the lower panel of figure~(\ref{fig:pos_P_single_traject}),
which shows the real part of this expression for a single stochastic run. Note that the \textit{averaged} rescaled $z$-components of the spins are always between $-1$ and $1$. 

Studying the lossy cavity, we are faced with typical problems that arise in positive $P$-representation simulations: individual stochastic trajectories show \textquotedblleft spikes\textquotedblright, as is
apparent in figure~(\ref{fig:pos_P_single_traject}). Such
spikes are a well-known problem in the context of
positive $P$-representation simulations~\cite{gardiner2004quantum,1464-4266-5-3-359,1751-8121-44-6-065305}. They indicate that the underlying
$P$-function is heavy-tailed. When the tails become so heavy that the second moment diverges, stochastic
averaging fails to converge beyond the time where spikes proliferate.
We analyze the spike statistics in the next section.
In extreme cases,
the positive $P$-function can even have non-vanishing mass at infinity, spoiling our derivation of the
Fokker-Planck equation, which relied on a partial integration and the dropping of surface terms.
In some cases, however, single trajectories can already predict much of the physics. In figure~\ref{fig:pos_P_single_traject}, we show such a characteristic
stochastic trajectory. We see the onset of self-trapping before the simulations break down. The dashed black curve in figure~\ref{fig:pos_P_single_traject}
shows the sum of both $z$-components of the two spins plus the total number of photons in the two cavities.
For the isolated system, this is a conserved quantity,
while for the open system this quantity is expected to smoothly decay. This is indeed what can be extracted from the plot, and as long as the dashed black curve is
continuous and smooth, the numerical simulations can be trusted. It is interesting to see from this plot that for some time before the simulations break down, single stochastic trajectories
undergo Rabi oscillations with the spin amplitude exceeding the classical allowed bound.
This is a clear signature of the simulations entering the quantum regime.
It was not possible for these parameters to carry out an ensemble average for long times without truncating divergent trajectories. We next consider a driven version of this system, which we can place into a steady state,
ameliorating such problems and demonstrating the power of the positive $P$-simulations.

\subsection{Driven dissipative Dicke dimer}
Having studied a strongly interacting quantum system  weakly coupled to the environment, we now
study the case of a strongly driven, dissipative system. As we shall see, strong drive and dissipation
will help to stabilize the positive $P$-representation simulations.
In the steady state of a driven system, autocorrelations quickly decay in time, and the spikes
are damped before having the opportunity to grow.
As the system will relax into a steady state, we are able to simulate long
times and even small spins.

We consider two coupled cavities each supporting an atom with spin $s$.
However, instead of filling the system initially with photons and letting them decay over time,
here we start with the cavities in the vacuum state and
coherently drive the left cavity, such that a steady state emerges. 
We choose a hopping rate $J=1$
(the other parameters are measured relative to $J$),
$\kappa = 20$ for both cavities, and set a coherent drive with amplitude $f=100/\sqrt{2}$.
In the absence of the second cavity, and for $g=0$, this would lead to a steady state photon number of $50$.
We vary the interaction strength
$g$ from $0$ to $10$. Our observable is the photon current. 

\paragraph*{\bf{Non-interacting limit}}
For $g=0$, the stochastic equations become deterministic and reduce to
($\alpha$'s are the expectation values of the annihilation operators in a coherent state)
\be
\dot{\alpha_1} & = & i J \alpha_2 - \kappa\alpha_1/2 + f \, , \\
\dot{\alpha_2} & = & i J \alpha_1 - \kappa\alpha_2/2 \, . \nonumber
\ee
While an analytic time-dependent solution exists, even more straightforwardly the steady state 
values can be derived by setting the time derivatives to zero, yielding
\be
\label{eq:equil_current}
j \;  = \;    \frac{16 \kappa f^2 J^2}{(\kappa^2 + 4J^2)^2} \; \approx  \;9.803 \,J .
\ee
We will use this result as a reference.
In the following we will consider the regime of finite $g$ and $s$.

\begin{figure}[h]
\begin{center}
\includegraphics[width = \linewidth]{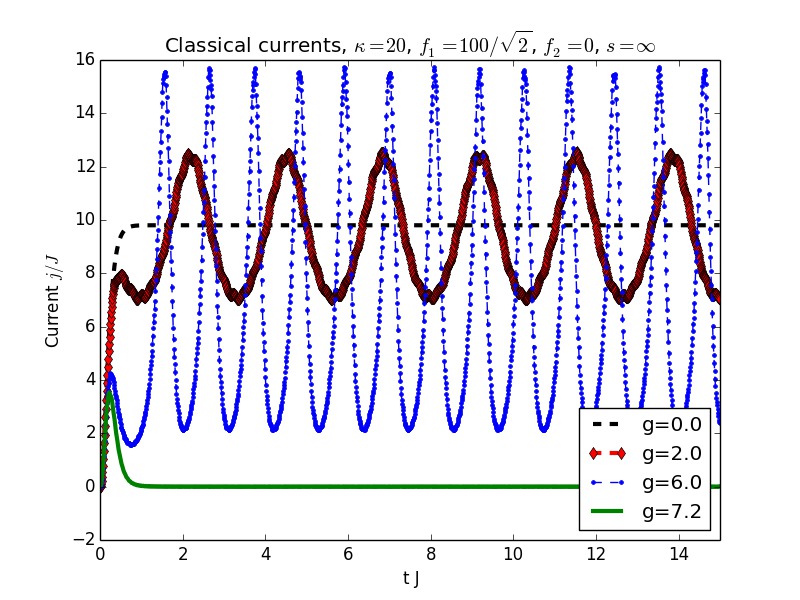}
\caption{ Classical, time-dependent current for $s=\infty$ and different values of $g$. Below the critical value of about $g_c \approx 7 J$, the current oscillates around its positive mean value. Above 
$g_c$, the current drops to zero. Quantum effects due to finite $s$ average out of these persistent oscillations and smooth the transition to a crossover, see also figures~\ref{fig:driven_quantum} and \ref{fig:g_plot}.
}
\label{fig:driven_classic}
\end{center}
\end{figure}

\paragraph*{\bf{Classical simulations, finite coupling $g$}}
In the limit of infinite spin, we simulated the deterministic equations numerically. 
Figure~\ref{fig:driven_classic} shows the time-dependent currents for different values of $g$. Below a critical value of $g_c \approx 7 J$, the current is seen to oscillate
around a positive mean value. Note that the current never changes sign. Above $g_c$, the current vanishes.
This delocalization-localization transition is what we want to simulate for finite spin,
using the positive $P$-representation.

\begin{figure}[h]
\begin{center}
\includegraphics[width = \linewidth]{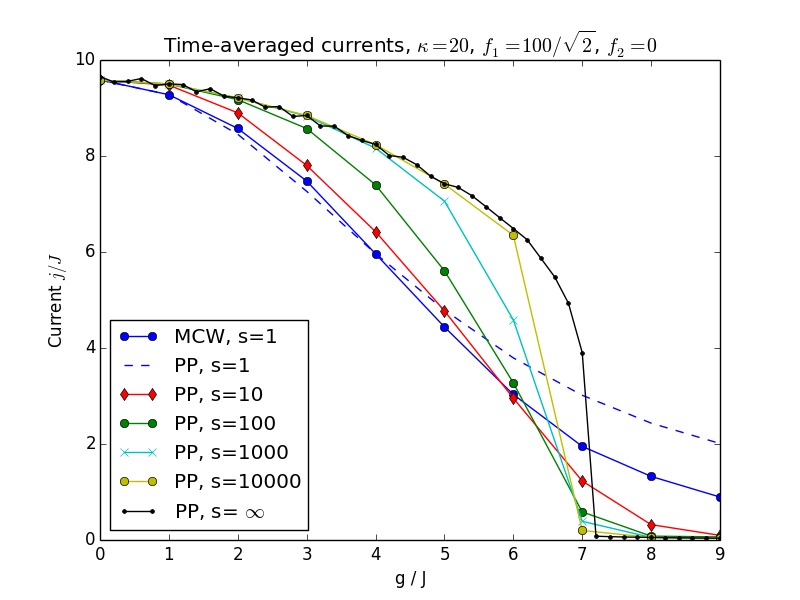}
\caption{Steady state currents $j$ as a function of the interaction strength $g$. Each data point results from an ensemble average 
over $6,000$ stochastic trajectories and a subsequent time average. 
While a sharp transition in the current is seen at infinite spin, finite values of $s$ turn this transition into a crossover. Deviations between MCW and PP simulations
grow close to the transition for $s=1$. As the sampling error  shrinks with larger spin size, we would expect better agreement for larger spins, which we cannot test due to the Hilbert space dimensionality
constraints of MCW.}
\label{fig:g_plot}
\end{center}
\end{figure}

\begin{figure}[h]
\begin{center}
\includegraphics[width = \linewidth]{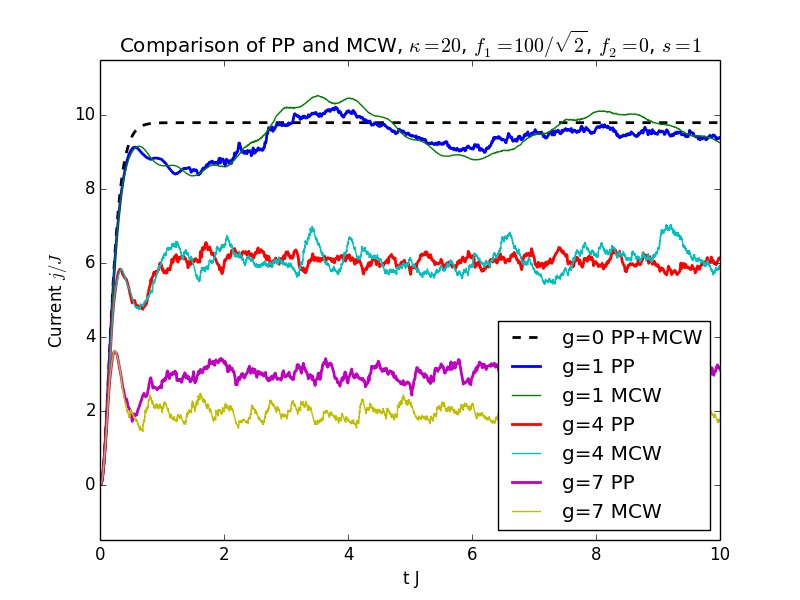}
\caption{Photon currents for spin $s=1$ (quantum case). We compare positive $P$ simulations (PP, averaged over 10000 trajectories) agains the Monte Carlo Wavefunction approach (MCW, averaged over 100 trajectories). 
A possible explanation for the systematic discrepancies at large $g$ is 
the fact that the photon currents show large statistical fluctuations in this regime that lead to negative photon currents for individual trajectories, see also figure~\ref{fig:histogram}.}
\label{fig:driven_quantum}
\end{center}
\end{figure}

\begin{figure}[h]
\begin{center}
\includegraphics[width = \linewidth]{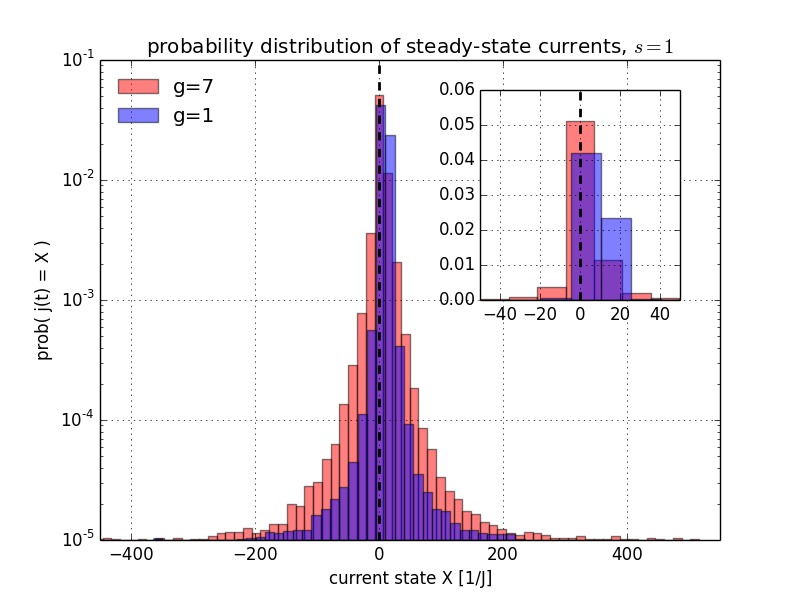}
\caption{Probability histogram of finding the current $j$ at a random time $t$ to be $X$ for a stochastic trajectory in steady state (log-scale).
The blue histogram in the foreground shows  $g=1$, where the stochastic fluctuations are much smaller than for $g=7$ (background, red).
Note that negative currents (flowing from the undriven, lossy cavity to the driven cavity) are unphysical, as are
negative photon  densities. 
When 
the statistical weight of such contributions is too large, the simulations lose their predictive power. The inset contains the same quantities presented on a non-logarithmic scale,
showing that the majority of trajectories have positive currents for $g=1$, but less so for $g=7$.}
\label{fig:histogram}
\end{center}
\end{figure}

\paragraph*{\bf{Quantum simulations}}
To study the behavior of the asymmetrically driven cavity in the quantum regime, we use the positive
$P$-representation,
scanning through all orders of magnitude of the spin $s$  in a range from $1$ to $10,000$.
We find that in the quantum case (finite $s$), the currents saturate to steady state values
that strongly depend on $g$. A strict phase transition only exists for $s=\infty$, but for large spins, the current is strongly suppressed above $g_c$.
This behavior is summarized in figure~\ref{fig:g_plot}. Here, the time averaged current is plotted as a function of $g$ for various spin sizes. 
Note that close to the transition, the statistical error
grows as the system becomes unstable due to the emergence of spikes.
Even for the case of $s=1$, a strong nonlinear dependence of the intercavity current on the interaction strength
$g$ is seen. This effect should be measurable in a circuit QED experiment.

We also compared our method against a numerical simulation based on the Monte Carlo Wavefunction
algorithm (MCW)~\cite{Molmer:93,PhysRevX.4.031043}.
This is an alternative approach based on an unraveling of the master equation, which allows one to simulate reasonable sized systems (the problem of an 
exponentially growing Hilbert space dimension still exists in this approach).
Figure~\ref{fig:driven_quantum} shows the outcome of a comparison of both methods.
In this figure, we plot the dynamics of the photon current as a function of time, starting with a
\textquotedblleft spin down\textquotedblright \ state and an empty dimer. The common parameters chosen are
$J=1,\kappa=20,f_1=100/\sqrt{2},f_2=0$, and we varied $g$ and $s$. We find good agreement in the time-dependent particle current for values of $g$ that 
are below the classical critical value of $g_c\approx 7 J $. For larger values of $g$, small
discrepancies appear. A possible explanation for this is the fact that 
deep in the quantum regime, individual stochastic trajectories may have negative particle numbers and negative currents. This behavior is also shown
in the histogram figure~\ref{fig:histogram}, which, at a given time $t$, counts the number of cases where the particle current is found at a given value. Upon normalization,
this can be interpreted as a probability distribution for the current. So long as
most of the mass sits in the positive range, positive $P$ simulations and MCW simulations agree reasonably well (here for $g=J$).
If much of the probability mass is in the forbidden region of negative currents, deviations become stronger and the positive $P$ simulations lose their validity.

\begin{figure}
\begin{center}
\includegraphics[width=\linewidth]{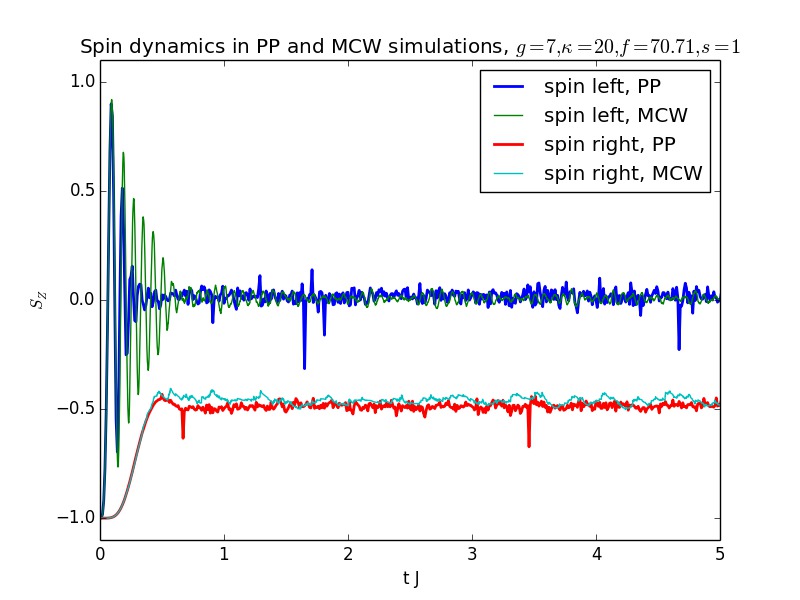}
\caption{Spin dynamics of the driven dimer, comparing positive $P$ simulations (PP, averaged over 10000 trajectories) against the Monte Carlo Wavefunction approach (MCW, averaged over 100 trajectories). The
$z$ components of the spins in the left (driven) cavity and in the right (undriven) cavity are shown as a function of time ($g=7J,s=1$). Note that the photon number in the undriven cavity is small, 
as the cavity loss rate $\kappa$  exceeds the incoming photon flow. Hence, the corresponding spin excitation is  small. 
In contrast, $S_z$ in the left cavity averages to zero due to rapid Rabi oscillations with the photon mode.
}
\label{fig:spin}
\end{center}
\end{figure}

A further comparison was carried out for the spin dynamics of the right and left cavity, as shown in
figure~\ref{fig:spin}, where also good agreement between positive $P$ stochastic simulations and MCW is obtained. 
While the undriven cavity saturates at a negative value for the $z$ component of the spin,  the driven cavity has an $S_z$ component that averages to zero. This can be understood as individual stochastic trajectories undergoing
Rabi oscillations with different relative phases, which averages out the $z$ component of the spin in the driven cavity.

This concludes our first application of the generalized positive $P$-representation as a numerical tool
for studying spin-boson systems.

\section{Summary and Conclusions}
\label{sec:summary}

We derived stochastic differential equations to model the nonequilibrium dynamics of systems involving bosons and quantum spins.
Our approach is based on a generalization of the positive $P$-representation using spin coherent states. This allows us to map a large class of Lindblad master equations
onto Fokker-Planck equations, following in a second step to a set of stochastic differential equations. 
Our approach can be applied to a variety of systems, including large networks.

Regarding computational efficiency, our approach scales linearly (instead of exponentially) with the number of network sites for nearest-neighbor couplings, and quadratically otherwise. We also note that,
in particular for problems involving coherent photons, we arrive at a much lower dimensional representation than in the usual Fock state representation.

We also modeled a dimer, each component consisting of a cavity coupled to a spin (of various sizes),
as a simple example.  
Individual stochastic trajectories were found to display heavy-tailed fluctuations, the so-called spikes.
Drive and dissipation reduce these fluctuations and bound the sampling variances. We compared our approach 
against the Monte Carlo Wavefunction method~\cite{Molmer:93}, and found good agreement. 
For the undriven, dissipative dimer,
we were able to qualitatively reproduce the super-exponential decay of the homodyne signal that has been observed in a recent circuit QED experiment~\cite{PhysRevX.4.031043}. 
We also studied the corresponding driven system
where we predicted a new phase transition in the inter-cavity current as a function of the on-site interaction strength. 

We plan to study larger systems than the dimer, i.e. large networks of cavities and spins.
For these systems, our method will compete very well with other existing methods
due to the favorable scaling properties of our approach.



\section*{Acknowledgements}
We would like to thank David Huse, Alexander Altland, Achim Rosch, Manas Kulkarni, and Matthias Troyer for stimulating discussions. 
Stephan Mandt acknowledges financial support from the NSF MRSEC program through the Princeton Center for 
Complex Materials Fellowship (DMR-0819860), and from the ICAM travel award (DMR-0844115).
The work of Darius Sadri, Hakan E. T\"{u}reci and Andrew A. Houck was supported by The Eric and Wendy Schmidt Transformative Technology Fund,
the US National Science Foundation through the Princeton Center for Complex Materials (DMR-0819860) and CAREER awards (Grant Nos. DMR-0953475 \& DMR-1151810),
the David and Lucile Packard Foundation,
and US Army Research Office grant  W911NF-11-1-0086.

\clearpage

\bibliography{paper}


\newpage

\begin{appendix}
\appendix

\section{Derivation of differential operator correspondences}
\label{appendix:spin_derivation}
In this section, we present and derive the correspondence between second-quantized operators and differential operators
that allowed us to derive the Fokker-Planck equation from the master equation. 
First, we use the following bosonic identities: 
\be
\label{photon_identities}
\hat{a} \hat{\Lambda} &=& \alpha \hat{\Lambda} \, , \\
\hat{\Lambda} \hat{a}  &=& \left(\partial_\beta+\alpha\right)\hat{\Lambda} \, , \nonumber \\
\hat{a}^\dagger \hat{\Lambda} &=&\left(\partial_\alpha+\beta\right)\hat{\Lambda} \, , \nonumber \\
\hat{\Lambda} \hat{a}^\dagger &=&\beta \hat{\Lambda} \, . \nonumber
\ee
Those identities are well-known and can be verified easily, see also Ref.~\cite{gardiner2004quantum}.
We likewise use expressions for the spin operators (proofs will be given below):
\be
\label{eq:spin-identities}
\hat{S}_+ \hat{\Lambda} &=& \left(\partial_z + \frac{2sw}{1+wz}\right) \hat{\Lambda} \, , \\
\hat{\Lambda} \hat{S}_+ &=& \left(-w^2\partial_w +\frac{2sw}{1+wz}\right)\hat{\Lambda} \, , \nonumber \\
\hat{S}_- \hat{\Lambda} &=& \left(-z^2\partial_z + \frac{2sz}{1+wz}\right)\hat{\Lambda} \, , \nonumber
\ee
\be
\hat{\Lambda} \hat{S}_- &=& \left(\partial_w +\frac{2sz}{1+wz}\right)\hat{\Lambda} \, , \nonumber \\
\hat{S}_z \hat{\Lambda} &=& \left(z\partial_z -s\frac{1-wz}{1+wz}\right)\hat{\Lambda} \, , \nonumber \\
\hat{\Lambda} \hat{S}_z &=& \left(w\partial_w -s\frac{1-wz}{1+wz}\right)\hat{\Lambda} \, \nonumber .
\ee

Those identities are very similar to
the identities used in Ref.~\cite{PhysRevLett.108.073601,1367-2630-14-7-073011} for the $Q-$representation,
but due to the doubling of degrees of freedom 
involved in the positive $P$-representation, the equations derived below slightly deviate from the latter ones. 
Also note that we use a different definition of the spin coherent states than in~\cite{PhysRevLett.108.073601,1367-2630-14-7-073011}, namely such that $z=0$
corresponds to a lowest weight state
(\textquotedblleft spin down\textquotedblright) instead of a highest weight state
(\textquotedblleft spin up\textquotedblright). This leads to a more
natural representation of the spin dissipators and the ground state (empty cavity without spin excitation).
 
First, we derive the spin identities presented previously in eq.~\eqref{eq:spin-identities}.
We begin with
\be
\hat{S}_+\hat{\Lambda} &\propto \hat{S}_+ e^{z\hat{S}_+}|s,-s\rangle =
\partial_z e^{z\hat{S}_+} |s,-s\rangle \, , \nonumber
\ee
where we did not yet respect the norm of $\hat{\Lambda}$.
Taking the latter into account yields the first identity in eq.~\eqref{eq:spin-identities}:
\be
\hat{S}_+\hat{\Lambda} =  \left(\partial_z + 2s\frac{w}{1+wz}\right)\hat{\Lambda} \, .
\ee
Deriving the second identity requires more work. We start with
\be
\hat{S}_-e^{z\hat{S}_+} |s,-s\rangle = [\hat{S}_-,e^{z\hat{S}_+}]|s,-s\rangle + e^{z\hat{S}_+}
\hat{S}_- |s,-s\rangle \nonumber \, .
\ee
The second term vanishes since the lowering operator annihilates the minimum weight states.
The exponential in the first term can be written as a taylor series, with terms of the following type:
\begin{widetext}
\begin{align}
\left[ \hat{S}_ -,(\hat{S}_+)^n \right] |s,-s\rangle
&= \sum_{l=0}^{n-1}(\hat{S}_+)^l
\underbrace{[\hat{S}_-,\hat{S}_+]}_{=-2\hat{S}_z}(\hat{S}_+)^{n-l-1} |s,-s\rangle
= \sum_{l=0}^{n-1}(\hat{S}_+)^l 2\left(-s+(n-l-1)\right)(\hat{S}_+)^{n-l-1} |s,-s\rangle \nonumber \\
&= (\hat{S}_+)^{n-1}\left(2ns-n (n-1)\right)\, |s,-s\rangle \, .
\end{align}
\end{widetext}
Hence,
\begin{widetext}
\begin{align}
[\hat{S}_-,e^{z \hat{S}_+}]  |s,-s\rangle
&= \sum_{n=0}^{\infty}\frac{z^n}{n!}[\hat{S}_-,(\hat{S}_+)^n]  |s,-s\rangle
= \left(2s\sum_{n=1}^{\infty}n  \frac{z^n}{n!} (\hat{S}_+)^{n-1} - \sum_{n=1}^{\infty}
\frac{z^n}{n!}n(n-1) (\hat{S}_+)^{n-1}  \right) |s,-s\rangle \nonumber \\
&= \left(2zs\sum_{n=0}^{\infty}n  \frac{z^n}{n!} (\hat{S}_+)^{n} -
z^2 \partial_z \sum_{n=0}^{\infty} \frac{z^n}{n!}(\hat{S}_+)^{n}  \right) |s,-s\rangle \n
&= (-z^2 \partial_z + 2 z s) \, e^{z \hat{S}_+}  |s,-s\rangle \, .
\end{align}
\end{widetext}
Again, taking derivatives with respect to the normalization into account yields 
\be
\hat{S}_- \hat{\Lambda} &=& \left[  -z^2 \partial_z + 2 z s + \left(z^2\partial_z\frac{1}{(1+wz)^{2s}} \right)(1+wz)^{2s}\right] \hat{\Lambda}\n
&=& \left(    -z^2 \partial_z +2s \frac{z+z^2w}{1+wz} - 2s\frac{z^2w}{1+wz}\right)\hat{\Lambda} \n
&=& \left(-z^2\partial_z + 2s\frac{z}{1+wz}\right)\hat{\Lambda}.
\ee

For the third equation in Eq.~(\ref{eq:spinidentities}), we start with
\be
\hat{S}_z e^{z \hat{S}_+}  |s,-s\rangle & = & \sum_{n=0}^{\infty}\frac{z^n}{n!} \hat{S}_z(\hat{S}_+)^n  |s,-s\rangle\\
& = & \sum_{n=0}^{\infty}\frac{z^n}{n!}(-s+n)(\hat{S}_+)^n  |s,-s\rangle \n
& = & (z\partial_z - s) e^{z \hat{S}_+}  |s,-s\rangle. \nonumber
\ee
As before, taking the derivatives on the normalization into account yields
\be
\hat{S}_z \hat{\Lambda} & = & \left(z\partial_z -s\frac{1-wz}{1+wz}\right)\hat{\Lambda}.
\ee
In order to derive the fourth, fifth and sixth identity in Eq.~(\ref{eq:spinidentities}), we use
\be
\hat{\Lambda} \hat{S}_+ & = \left(\hat{S}_- \hat{\Lambda}^\dagger\right)^\dagger&= \left(-w^2\partial_w +2s\frac{w}{1+wz}\right)\hat{\Lambda} \n
\hat{\Lambda} \hat{S}_- & =  \left(\hat{S}_+ \hat{\Lambda}^\dagger\right)^\dagger&=\left(\partial_w +2s\frac{z}{1+wz}\right)\hat{\Lambda} \n
\hat{\Lambda} \hat{S}_z& =  \left(\hat{S}_z \hat{\Lambda}^\dagger\right)^\dagger&=\left(w\partial_w -s\frac{1-wz}{1+wz}\right)\hat{\Lambda}. \nonumber
\ee
This concludes the proof of Eq.~(\ref{eq:spinidentities}).

\section{Fokker Planck equation I\\ (Hamiltonian contribution) }

We are now going to derive the Fokker-Planck equation term by term, starting from the 
Hamiltonian contributions. First, consider the interaction Hamiltonian.
\begin{widetext}
\be
[\hat{a}^\dagger \hat{S}_-,\hat{\Lambda}]  =   \left[ (\partial_\alpha + \beta)(-z^2\partial_z + 2s\frac{z}{1+wz}) -   \beta(\partial_w+ 2s\frac{z}{1+wz})\right] \hat{\Lambda} 
  =  \left( -z^2\partial_\alpha \partial_z +2s\frac{z}{1+wz}\partial_\alpha - \beta \partial_w - z^2\beta\partial_z \right)\hat{\Lambda}, \n
\left[\hat{a} \hat{S}_+,\hat{\Lambda} \right] =   \left[ \alpha(\partial_z + 2s\frac{w}{1+wz}) -   (\partial_\beta+\alpha)(-w^2 \partial_w+ 2s\frac{w}{1+wz})\right] \hat{\Lambda} 
  =  \left( w^2\partial_\beta\partial_w -2s\frac{w}{1+wz}\partial_\beta  + \alpha \partial_z + \alpha w^2 \partial_w \right)\hat{\Lambda}, \n
\Longrightarrow \left[\hat{a}^\dagger \hat{S}_- + \hat{a} \hat{S}_+ ,\hat{\Lambda} \right]=  \left[(w^2 \partial_w \partial_\beta - z^2 \partial_\alpha \partial_z) + (\alpha-\beta z^2)\partial_z - (\beta - \alpha w^2)\partial_w + 2s\frac{1}{1+wz}(z \partial_\alpha - w \partial_\beta)\right] \hat{\Lambda}.\n
\ee
\end{widetext}
The operators associated with the cavity frequency map according to
\be
[\hat{a}^\dagger \hat{a},\hat{\Lambda}] & = & [(\partial_\alpha + \beta)\alpha - (\partial_\beta + \alpha)\beta] \hat{\Lambda} \n
& = & [ \alpha \partial_\alpha- \beta\partial_\beta ] \hat{\Lambda}.
\ee
The spin frequency term yields
\be
\left[ \hat{S}_z, \hat{\Lambda} \right] & = & [(z\partial_z -  s\frac{1-wz}{1+wz}) - w\partial_w +  s\frac{1-wz}{1+wz}) ] \hat{\Lambda} \n
&  =& (z\partial_z - w\partial_w)\hat{\Lambda}.
\ee
Operators associated with a coherent drive result in 
\be
\left[ \hat{a}^\dagger - \hat{a},\hat{\Lambda} \right] & = & [(\partial_\alpha + \beta - \alpha) - (\partial_\beta + \beta - \alpha)]\hat{\Lambda} \n
& = & (\partial_\alpha - \partial_\beta)\hat{\Lambda}.
\ee
The kinetic energy term results in
\be
-iJ_{ij}[ (\hat{a}_i^\dagger \hat{a}_j + \hat{a}_j^\dagger \hat{a}_i),\hat{\Lambda} ] & = & -iJ_{ij}[(\partial_{\alpha_i}+\beta_i)\alpha_j + (\partial_{\alpha_j}+\beta_j)\alpha_i \n
    & & - (\partial_{\beta_j}+\alpha_j)\beta_i - (\partial_{\beta_i}+\alpha_i)\beta_j ]\hat{\Lambda} \n
& = & [-iJ_{ij} \alpha_j\partial_{\alpha_i} -iJ_{ij} \alpha_i\partial_{\alpha_j} \n
  & & +  iJ_{ij} \beta_j\partial_{\beta_i} + iJ_{ij} \beta_i\partial_{\beta_j}]\hat{\Lambda}.
\ee
This concludes the Hamiltonian contributions to the  Fokker-Planck equation.

\section{Fokker Planck equation II \\ (dissipators) }

\paragraph{\bf{Photon dissipators}}
First, we will calculate the dissipators of the photon fields, which are given by
\be
{\cal L}^{a}_{\rm out}[\hat{\Lambda}] & = & \frac{\kappa}{2}(\bar{n}+1)\left(2 \hat{a} \hat{\Lambda} \hat{a}^\dagger - \hat{a}^\dagger \hat{a} \hat{\Lambda} - \hat{\Lambda} \hat{a}^\dagger \hat{a}\right) \\
& = & \frac{\kappa}{2}(\bar{n}+1)\left[ 2 \alpha \beta - \alpha (\partial_\alpha + \beta) - \beta(\partial_\beta + \alpha)\right] \hat{\Lambda} \n
& = & \frac{\kappa}{2}(\bar{n}+1)\left[ - \alpha \partial_\alpha - \beta \partial_\beta\right] \hat{\Lambda} \nonumber.
\ee
Similarly, we find for the ingoing term,
\be
& & {\cal L}^{a}_{\rm out}[\hat{\Lambda}] \\
 & = & \frac{\kappa}{2}\bar{n}\left(2 \hat{a}^\dagger \hat{\Lambda} \hat{a}- \hat{a} \hat{a}^\dagger \hat{\Lambda} - \hat{\Lambda} \hat{a} \hat{a}^\dagger \right) \n
& = & \frac{\kappa}{2}\bar{n}\left[ 2 (\partial_\alpha + \beta) (\partial_\beta + \alpha) - (\partial_\alpha + \beta)\cdot \alpha - (\partial_\beta + \alpha)\cdot \beta \right] \hat{\Lambda} \n
& = & \frac{\kappa}{2}\bar{n}\left[ 2 \partial_\alpha \partial_\beta + 2 \beta \partial_\beta  + 2 \partial_\alpha \cdot \alpha  - \partial_\alpha \cdot \alpha - \partial_\beta \cdot \beta \right] \hat{\Lambda} \n
& = & \frac{\kappa}{2}\bar{n}\left[ 2 \partial_\alpha \partial_\beta + \beta \partial_\beta + \alpha \partial_\alpha \right] \hat{\Lambda} \nonumber.
\ee
Hence,
\be
{\cal L}^{a}_{\rm}[\hat{\Lambda}] & = & {\cal L}^{a}_{\rm out}[\hat{\Lambda}] + {\cal L}^{a}_{\rm in}[\hat{\Lambda}]  \\[.1cm]
& = & \frac{\kappa}{2} \left( - \alpha \partial_\alpha - \beta \partial_\beta \right) + 2 \bar{n} \,\partial_\alpha \partial_\beta) \hat{\Lambda}. \nonumber
\ee
Interestingly, note that  there is no noise term for $\bar{n}=0$. In the positive $P$-representation
at zero temperature, all noise comes from quantum fluctuations, and its strength depends on $g$ as opposed to $\kappa$.

\paragraph{\bf{Spin dissipators}}
In contrast to the dissipators for the photon field, calculating the spin dissipators, ~\eq{eq:spindiss}, is much more work.
Again, we distinguish between ``in'' dissipators (existing only at finite temperature), and ``out'' dissipators. Let us calculate them
term by term, using~\eq{eq:spinidentities}:
\be
& & {\cal L}^{S}_{\rm out}[\hat{\Lambda}] \n
& = & \frac{\gamma}{2}(\bar{n}+1)\left( 2 \hat{S}_- \hat{\Lambda} \hat{S}_+    -  \hat{S}_+\hat{S}_-\hat{\Lambda} -   \hat{\Lambda} \hat{S}_+\hat{S}_- \right) \n
 & = &  \frac{\gamma}{2}(\bar{n}+1)\left[  2(-z^2 \partial_z + 2s\frac{z}{1+wz})(-w^2 \partial_w + 2s\frac{w}{1+wz})\right. \n
&  & -(-z^2 \partial_z + 2s \frac{z}{1+wz})(\partial_z + 2s \frac{w}{1+wz}) \n
  & & -\left. (-w^2 \partial_w + 2s \frac{w}{1+wz})(\partial_w + 2s \frac{z}{1+wz})  \right] \hat{\Lambda}.
\ee
Denoting Lindblad terms containing first and second order differential operators as $ {\cal L}^{S\, (1)}_{\rm out}$ and  $ {\cal L}^{S\, (2)}_{\rm out}$,
respectively, we find
\be
 {\cal L}^{S\, (2)}_{\rm out} & = & \frac{\gamma}{2}(\bar{n}+1) \left[ 2z^2w^2 \partial_z \partial_w + z^2 \partial_z^2 + w^2 \partial_w^2\right] \hat{\Lambda} ,\n
 {\cal L}^{S\, (1)}_{\rm out} & = & \frac{\gamma}{2}(\bar{n}+1) \left[ -2z^2 \partial_z \cdot 2s\frac{w}{1+wz} - 4s\frac{z}{1+wz} w^2 \partial_w \right. \n
& & + z^2 \partial_z \cdot 2s\frac{w}{1+wz} - 2s\frac{z}{1+wz}\partial_z \n
& & + \left. w^2 \partial_w \cdot 2s \frac{z}{1+wz} - 2s \frac{w}{1+wz}\partial_w \right] \hat{\Lambda}.
\ee
Using the identity $\partial_z \cdot 2s\frac{w}{1+w} = 2s\frac{w}{1+w} \partial_z - 2s\frac{w^2}{(1+w)^2}$ 
(and the same identity for $z$ and $w$ interchanged) results in
\be
 {\cal L}^{S\, (1)}_{\rm out} & = & -\frac{\gamma}{2}(\bar{n}+1) 2s \left[ \frac{z+z^2w}{1+wz}\partial_z + \frac{w+w^2z}{1+wz}\partial_w \right]\hat{\Lambda} \n
 & = &  -\frac{\gamma}{2}(\bar{n}+1) 2s (z\partial_z + w \partial_w).
\ee
Now, let's consider the ``in'' term,
\be
{\cal L}^{S}_{\rm in}[\hat{\Lambda}] 
& = & \frac{\gamma}{2}\bar{n}\left( 2 S+ \hat{\Lambda} \hat{S}_-    -  \hat{S}_-\hat{S}_+\hat{\Lambda} -   \hat{\Lambda} \hat{S}_-\hat{S}_+ \right) \\
 & = &  \frac{\gamma}{2}\bar{n}\left[  2(\partial_z + 2s\frac{w}{1+wz})(\partial_w + 2s\frac{z}{1+wz})\right. \n
&  & -(\partial_z + 2s \frac{w}{1+wz})(-z^2\partial_z + 2s \frac{z}{1+wz}) \n
  & & -\left. (\partial_w + 2s \frac{z}{1+wz})(-w^2 \partial_w + 2s \frac{w}{1+wz})  \right] \hat{\Lambda}. \nonumber
\ee
Again, collecting second and first order differential operators, and doing a similar calculation as above for the latter results in
\be
 {\cal L}^{S\, (2)}_{\rm in} & = & \frac{\gamma}{2}\bar{n} \left[ \partial_z \partial_w + z^2 \partial_z^2 + w^2 \partial_w^2\right] \hat{\Lambda}, \\
 {\cal L}^{S\, (1)}_{\rm in} & = & \frac{\gamma}{2}\bar{n} \left[ 2(s+1) z \partial_z + 2(s+1) w \partial_w \right] \hat{\Lambda}. \nonumber
\ee
The full Lindblad dissipators, containing first and second order differentials, result as a sum ingoing and outgoing terms,
\be
 {\cal L}^{S\, (2)} & = &  \left[-\frac{\gamma}{2}(2\bar{n}+1)(z^2 \partial_z^2 + w^2 \partial_w^2)\right. \n
& & + \left.\gamma(\bar{n} + (\bar{n}+1)z^2 w^2)\partial_z \partial_w \right] \hat{\Lambda},\n
 {\cal L}^{S\, (1)} & = &  \left[- \frac{\gamma (\bar{n}+1)}{2} 2s (z\partial_z + w \partial_w) \right. \n
& & \left. \frac{\gamma \bar{n}}{2} (2s+1) (z\partial_z + w \partial_w)  \right] \hat{\Lambda} \n
&= &\left[ \gamma(-(\bar{n}+1)s z + \bar{n} \frac{2s+1}{2} z) \partial_z + (z \leftrightarrow w)\right] \hat{\Lambda}. \nonumber  
\ee

\section{Operator expectation values}
\label{appendix:spin_expectation_values}
We already indicated that the positive $P$-function allows to calculate expectation values of bosonic field operators.
Similarly, also spin expectation values can be calculated and arbitrary mixed expectation values, as we will show now.
To this end, we need the following identities that are straightforward consequences of the definitions of spin coherent states:
\be
\langle z| S_x | w\rangle & = & s\frac{z + w}{1 + zw}, \\
\langle z| S_y | w\rangle & = &\frac{s}{i}\frac{z - w}{1 + zw}, \\
\langle z| \hat{S}_z | w\rangle & = & s\frac{1 - zw}{1 + zw}. 
\ee
An operator expectation value involving e.g. $\hat{S}_z$ would therefore amount to calculating
\be
\langle \hat{S}_z \rangle & =&  \int d\balpha P(\balpha,t)\, {\rm Tr} [\hat{S}_z \hat{\Lambda}(\balpha)]\\
& = & \int d\balpha \,P(\balpha,t)  \, s\frac{1 - zw}{1 + zw}
\ee
and so on.
\section{numerical regularization}
When numerically simulating the stochastic differential equations, certain regularizations have to be applied to guarantee numerical stability.
First, note that in the spin coherent state representation, the lowest weight state (``spin down'') corresponds to $z=0$, while the highest weight (``spin up'')
corresponds to $z=\infty$. Hence, a rigorous ``spin up'' state can only be approximated in our representation. If the initial state is prepared for $z=0$ and the photon field
is coherent, Rabi oscillations will typically dynamically drive the spin to a highest weight state, leading to a breakdown of the numerics without regularization. 

We use the following tricks to avoid this problem. First, we found that numerical stability is enhanced when the spin coherent state slightly deviates initially from 
$z=0$ by e.g. initializing $z=\epsilon_1 + i \epsilon_2$, and  $w=\epsilon_1 - i \epsilon_2$ where $0<\epsilon_{1},\epsilon_2<10^{-5}$ at time $t=0$. This trick is not necessary in the 
presence of thermal or quantum noise, which we found to enhance stability in this respect. More importantly,
we add a regularizing term to the stochastic differential equations for $z$ and $w$. To be precise, we replace the stochastic differential equations~(\ref{eq:stoch_diff_general})
by
\be
d\balpha &=& A(\balpha) dt \, +\,  \xi(\balpha,t) - \,R(z,w)dt
\ee
where
\be
R(z,w) &=& (0,0,r(z),r(w))^T,\\
r(x) & = & (e^{\epsilon |x|^2}-1) \, x/|x|,\quad\epsilon = 10^{-8}.
\ee
Hence, we add a ``restoring force'' which grows exponentially at very large radii in the complex plane for $z$ and $w$. Under the stereographic mapping,
 this region on the complex plane corresponds to a very tiny ``polar region'' around the Bloch sphere's north pole (highest weight state).
Strictly speaking, the regularization term violates certain symmetries such as the conservation of total excitations per cavity, 
but we carefully checked that those effects are extremely small and negligible due to the smallness of $\epsilon$.

\section{Mapping to spherical coordinates}
In the absence of quantum noise, i.e. in the scaling limit of $s\rightarrow \infty$, our positive
$P$-representation
becomes equivalent to the P representation. To see this, note that the deterministic equations~\ref{eq:determin_A}
have the property that for initial conditions $\alpha(0) = \beta^*(0)$ and $z(0) = w^*(0)$,
the pairs $\alpha,\beta$ and $z,w$ stay complex conjugates for all times. Note that the thermal noises that act on
$\alpha$ and $\beta$ are also complex conjugates by construction. Therefore, the equations for $\beta$ and $w$ are
redundant in this limit, and it is enough to simulate the dynamics of $\alpha$ and $z$. This corresponds to the P representation.

Let us consider the P representation. It turns out that the following coordinate transformation yields a set of stochastic differential equations
with a better numerical stability.  A closely related transformation has been carried out
in~\cite{PhysRevLett.108.073601,1367-2630-14-7-073011}, but in contrast to the latter, we keep
the variable $\alpha$. We consider the inverse stereographic projection, mapping the spin field 
back on the sphere,
\be
z = \sqrt{\frac{1-c}{1+c}} e^{i\phi}
\ee
where $\phi \in {\mathbb R}$ and $c \in [-1,1]$.
We only transform the spin part and leave the equations for the photon field $\alpha$ unchanged.
This transformation results in a new stochastic differential equation of the form
\be
d\alpha & = & A_1(\alpha,\phi,c)dt + \sqrt{\bar{n}\kappa}\, (dW_1 + i dW_2)/\sqrt{2 } \label{eq:SDE_spherical} \\
d \phi & = & (\omega_z + g \frac{c}{\sqrt{1-c^2}} (\alpha^* e^{i \phi} + \alpha e^{-i\phi}) ) dt \n
d c & = & g \sqrt{1-c^2} (i\alpha^* e^{i \phi} - i\alpha e^{-i\phi})\,dt \nonumber
\ee
The function $A_1$ is given by
\be
A_1(\alpha,\phi,c) & = & (i \omega_c  - \kappa/2)\alpha - i J \bar{\alpha} + f + i g \sqrt{1-c^2} e^{i\phi}. \n
\ee
Here, we focussed on a single cavity, and $\bar{\alpha}$ is the field in the other cavity. 
We used this set of equations when simulating the finite temperature dynamics of the system in
the scaling limit of infinite spin.


\end{appendix}

\end{document}